\documentclass[letterpaper, 10 pt, conference]{ieeeconf}  

\IEEEoverridecommandlockouts                              

\usepackage{graphicx}
\usepackage{tikz}
\usepackage{color}
\usepackage{amsmath}
\usepackage{subfig}
\usepackage{amssymb}

\title{\LARGE \bf
Reverberant Sound Localization with a Robot Head\\Based on Direct-Path Relative Transfer Function
}

\author{Xiaofei Li$^{\dag}$, Laurent Girin$^{\dag,\ddag,\#}$, Fabien Badeig$^{\dag}$ and Radu Horaud$^{\dag}$ \\
	$^{\dag}$INRIA Grenoble Rhone-Alpes, $^{\ddag}$GIPSA-LAB,  $^{\#}$Univ. Grenoble Alpes
\thanks{This research has received funding from the EU-FP7 STREP project EARS (\#609465).}
}

\begin{document}
\bibliographystyle{ieeetr}

\maketitle
\thispagestyle{empty}
\pagestyle{empty}

\begin{abstract}
This paper addresses the problem of sound-source localization (SSL) with a robot head, which remains a challenge in real-world environments. In particular we are interested in locating speech sources, as they are of high interest for human-robot interaction. The microphone-pair response corresponding to the direct-path sound propagation is a function of the source direction. In practice, this response is contaminated by noise and reverberations. The direct-path relative transfer function (DP-RTF) is defined as the ratio between the direct-path acoustic transfer function (ATF) of the two microphones, and it is an important feature for SSL. We propose a method to estimate the DP-RTF from noisy and reverberant signals in the short-time Fourier transform (STFT) domain. First, the convolutive transfer function (CTF) approximation is adopted to accurately represent the impulse response of the microphone array, and the first coefficient of the CTF is mainly composed of the direct-path ATF. At each frequency, the frame-wise speech auto- and cross-power spectral density (PSD) are obtained by spectral subtraction. Then a set of linear equations is constructed by the speech auto- and cross-PSD of multiple frames, in which the DP-RTF is an unknown variable, and is estimated by solving the equations. Finally, the estimated DP-RTFs are concatenated across frequencies and used as a feature vector for SSL. Experiments with a robot, placed in various reverberant environments, show that the proposed method outperforms two state-of-the-art methods. 
\end{abstract}

\section{Introduction}

\begin{figure}[t]
\centering
{\includegraphics[width=0.45\textwidth]{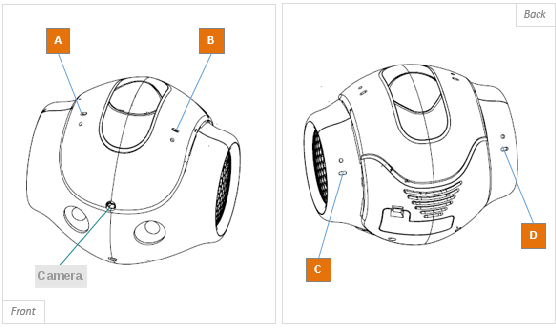}}
\caption{The version~5 of the NAO head has four microphones, namely A, B, C, and D. This robot-head configuration is used in our experiments to illustrate the proposed SSL method.} 
\label{fignao}
\end{figure}

Sound source localization (SSL) is a crucial methodology for robot audition. 
This paper addresses the problem of real-world SSL using a microphone array embedded into a robot head. 
The NAO robot (version 5) is used in this paper, whose head and its four embedded microphones are shown on Fig.~\ref{fignao}.

Microphone-array processing SSL techniques are widely adopted for robot audition, e.g., \cite{argentieri2007,nakamura2012,valin2004,valin2007,sasaki2013,gomez2015}. 
These techniques generally need a large number of microphones and high computational cost. The time difference of arrival (TDOA) techniques \cite{badali2009,alameda2014} are suitable if fewer microphones are available, however 
they are generally applied to a free-field setup, in which the TDOA is frequency-independent.
We address SSL in the more general case, namely when the source-to-sensor sound propagation is affected by the robot's head and torso, e.g., binaural audition \cite{hornstein2006,raspaud2010}, as well as by the room acoustics \cite{deleforge2015a}, and these effects are frequency-dependent \cite{blauert1997}. 

As shown in Fig. \ref{fignao}, four microphones are embedded in NAO's head. The two most discriminative microphone pairs in terms of SSL, i.e., the two cross microphone pairs (A-C and B-D) are used in this paper.  
The acoustic features are extracted separately from these two microphone pairs, and then these \textit{pairwise features} are combined together.
Two interaural cues, the interaural time (or phase) difference (ITD or IPD) and  the interaural level difference (ILD), are widely used for SSL. 
When computed using the STFT, the ILD and IPD correspond to the magnitude and phase of a two-channel \textit{relative transfer function} (RTF), which is the ratio between the ATFs of  the two microphones \cite{gannot2001}. 
The interaural cues, or equivalently the two-channel RTF, that correspond to the direct-path sound propagation are a function of the source direction, which is to be estimated from noisy and reverberant sensor signals, as they are available in a real environment.

Techniques have been proposed to identify the RTF in noisy environments, such as a speech non-stationary estimator \cite{gannot2001},  an RTF identification method based on speech presence probability and spectral subtraction \cite{cohen2004}, and
an RTF estimator based on segmental PSD matrix subtraction \cite{mine2015assp}. In these RTF estimators, the multiplicative transfer function (MTF) approximation \cite{avargel2007spl} is assumed. This approximation is justified only when the length of the room impulse response is shorter than the length of the STFT window, which is rarely the case in realistic acoustic setups. 
Moreover, the RTF estimated above is the ratio between two ATFs that include the reverberations and hence it is poorly suitable for SSL in echoic environments.

Techniques have been proposed to extract the interaural cues that correspond to the direct-path sound propagation, 
e.g. based on the detection of time frames with less reverberations. The precedence effect \cite{litovsky1999} is widely modeled for SSL \cite{bechler2005,hummersone2013}, which relies on the principle that the onset frame is dominated by the direct-path wavefront \cite{may2011probabilistic,woodruff2012binaural}. 
In the STFT domain, the coherence test \cite{mohan2008} and the direct-path dominance test \cite{nadiri2014} are proposed to detect the frames dominated by one active source (namely only the direct-path propagation), from which reliable localization cues are estimated.
However, in practice, there are always reflection components in the frames selected by these algorithms due to the inaccurate model or an improper decision threshold. 

In this paper we propose a direct-path RTF estimator suitable for the localization of a single speech source in the real world. We build on the crossband filter proposed in \cite{avargel2007} (actually a simplified CTF approximation proposed in \cite{talmon2009}) for system identification. 
This filter accurately characterizes the impulse response in the STFT domain by a convolutive transfer function instead of the MTF approximation. The first coefficient of the CTF at different frequencies represents the STFT of the first segment of the channel impulse response, which is composed of the impulse response of the direct-path propagation and possibly a few reflections. 
Therefore, we refer to the first coefficient of the CTF as the direct-path ATF, and the ratio between the coefficients from the two channels as the direct-path RTF (DP-RTF).
For the noise-free case, inspired by \cite{benesty2000}, based on the relation of the CTFs between the two channels, we construct a set of linear equations using the auto- and cross-power spectral density (PSD) of the speech signal received by the microphones. 

At each frequency, the DP-RTF is the unknown variable of the linear equations, and can be estimated from these equations using the least square estimator. 
However, in practice, the sensor signals are always contaminated by noise. The speech PSD constructing the linear equations can be obtained by subtracting the noise PSD from the sensor signal PSD. 
Finally, the estimated DP-RTFs are concatenated over microphone pairs and frequencies, and mapped to the source direction space using the probabilistic piecewise affine mapping model \cite{deleforge2015b}. 
Experiments, conducted in various real-world environments, show the effectiveness of the proposed method.

The remainder of this paper is organized as follows. Section~\ref{secformu} formulates the sensor signals based on the crossband filter. Section~\ref{secdprtf} presents the DP-RTF estimator. 
In Section~\ref{secssl}, the SSL algorithm based on the probabilistic piecewise affine mapping model is described.  Experimental results are presented in Section~\ref{secexp}, and Section~\ref{seccon} draws some conclusions.

\section{Signal Formulation Based on Crossband Filter}
\label{secformu}
In this work, we process microphone pairs separately. Thence, without loss of generality, only the sensor signals of one microphone pair are defined in this section, analyzed in section \ref{secdprtf}, and the acoustic features of several microphone pairs will be combined for SSL in Section \ref{secssl}.

Let us consider a non-stationary source signal, e.g., a speech source $s(n)$ in the time domain. In a noise-free environment, the microphone-pair signals are
\begin{align}\label{xn}
 x(n)=a(n)*s(n), \quad  y(n)=b(n)*s(n), 
\end{align}
where $*$ denotes convolution, $a(n)$ and $b(n)$ are the room impulse responses from the source to the first and second microphone, respectively. 
Let $T$ denote the length of $a(n)$ and $b(n)$.
Applying the STFT, based on the MTF approximation, microphone signal $x(n)$ is approximated in the time-frequency (TF) domain as $x_{p,k}=s_{p,k}a_k$, 
where $x_{p,k}$ and $s_{p,k}$ are the STFT of the corresponding signals, $p$ and $k$ are the indexes of time frame and frequency bin, respectively. Let $N$ denote the length of the STFT window (frame). 
This MTF approximation is only valid when the impulse response length $T$ is lower than $N$. 
For a non-stationary acoustic signal, such as speech, a small length $N$ (around 20 ms) is typically chosen to assume `local' stationarity, i.e. in each frame. 
Therefore the MTF approximation is questionable in a, possibly strongly, reverberant environment with a long room impulse response.

To address this problem, the crossband filter was introduced in \cite{avargel2007} to  represent a linear system in the STFT domain more accurately. Let $\tilde{\omega}(n)$ and $\omega(n)$ denote the analysis and synthesis STFT windows respectively, and let $L$ denote the frame step.  
The crossband filter model consists of representing the STFT coefficient $x_{p,k}$ as a summation of multiple convolutions across frequency bands.
A CTF approximation is further introduced in \cite{talmon2009} to simplify the analysis, i.e. using only band-to-band filters as
\begin{align}\label{xpk3}
 x_{p,k} &= \sum_{p'=0}^{Q_k-1} s_{p-p',k}a_{p',k}= s_{p,k}*a_{p,k},
\end{align}
where convolution is applied to the time variable $p$. The frequency dependent CTF length $Q_k$ is related to the reverberation at the $k$th frequency band, which will be discussed in section \ref{secexp}. 
The TF-domain impulse response $a_{p',k}$ is related to the time-domain impulse response $a(n)$ by:
\begin{align}\label{hp}
a_{p',k}={a(n)*\zeta_{k}(n)}|_{n=p'L},
\end{align}
which represents the convolution with respect to the time index $n$ evaluated at frame steps, with
\begin{align}\label{phik}
\zeta_{k}(n) = e^{j\frac{2\pi}{N}kn}\sum_m\tilde{\omega}(m)\omega(n+m).
\end{align}
In the next section, the CTF formalism is used to extract the impulse response of the direct-path propagation.

%

%

\section{Direct-Path Relative Transfer Function}
\label{secdprtf}
\subsection{Definition of DP-ATF and DP-RTF Based on CTF}
\label{dpatf}

In the CTF approximation (\ref{xpk3}), using (\ref{hp}) and (\ref{phik}) at $p'=0$, the first coefficient of $a_{p',k}$ can be derived as 
\begin{align}
 a_{0,k} = {a(n)*\zeta_{k,k}(n)}|_{n=0} = \sum\nolimits_{t=0}^{N-1} a(t)\nu(t)e^{-j\frac{2\pi}{N}kt},
\end{align}
where  
\begin{equation}
 \nu(t)=
 \begin{cases} \sum_{m=0}^{N}\tilde{\omega}(m)\omega(m-t)  & \mbox{if } 1-N\le t\le N-1, \\
  0, & \mbox{otherwise.}
 \end{cases} \nonumber
\end{equation}
Therefore, $a_{0,k}$ can be interpreted as the $k$-th Fourier coefficient of the impulse response segment $a(n)|_{n=0}^{N-1}$ (windowed by $\nu(t)|_{n=0}^{N-1}$).
In the sense of transfer function identification, without loss of generality, we assume that the room impulse response $a(n)$ begins with the impulse response of the direct-path sound propagation.    
If the frame length $N$ is properly chosen, $a(n)|_{n=0}^{N-1}$ is composed of the impulse responses of the direct-path propagation and a few reflections. 
Particularly, if the initial time delay gap (ITDG) is large compared to the frame length $N$, $a(n)|_{n=0}^{N-1}$ is mainly composed of the direct-path impulse response. Thence we refer to $a_{0,k}$ as the direct-path ATF.

Similarly, the CTF approximation of $y_{p,k}$ is written as
\begin{align}\label{ypk3}
 y_{p,k} = s_{p,k}*b_{p,k},
\end{align}
and $b_{0,k}$ is assumed to represent the direct-path ATF from the source to the second microphone. By definition, DP-RTF is given by: $\frac{b_{0,k}}{a_{0,k}}$.
Let us remind that this DP-RTF is a relevant cue for SSL.

\subsection{DP-RTF Estimation}
\label{sec22}

Since both channels are assumed to follow the CTF model, we can write:
\begin{align}\label{xyha}
 x_{p,k}*b_{p,k}=s_{p,k}*a_{p,k}*b_{p,k}=y_{p,k}*a_{p,k}.
\end{align}
In \cite{benesty2000}, this relation is proposed in time domain for TDOA estimation. 
Eq.(\ref{xyha}) can be written in vector form as
\begin{align}\label{mxa}
 \mathbf{x}_{p,k}^{\top} \mathbf{b}_k = \mathbf{y}_{p,k}^{\top} \mathbf{a}_k
\end{align}
where $^{\top}$ denotes vector or matrix transpose, and 
\begin{align}
 &\mathbf{x}_{p,k} = [x_{p,k},x_{p-1,k},\dots,x_{p-Q_k+1,k}]^{\top}, \nonumber \\
 &\mathbf{y}_{p,k} = [y_{p,k},y_{p-1,k},\dots,y_{p-Q_k+1,k}]^{\top}, \nonumber \\
 &\mathbf{b}_k = [b_{0,k},b_{1,k},\dots,b_{Q_k-1,k}]^{\top}, \nonumber \\
 &\mathbf{a}_k = [a_{0,k},a_{1,k},\dots,a_{Q_k-1,k}]^{\top}.
\end{align}
Dividing both sides of (\ref{mxa}) by $a_{0,k}$ and reorganizing the terms, we can write:
 \begin{align}\label{zpk}
 y_{p,k} = \mathbf{z}_{p,k}^{\top} \mathbf{g}_k,
 \end{align}
where
\begin{align}\label{zg}
 &\mathbf{z}_{p,k}=[x_{p,k},\dots,x_{p-Q_k+1,k},y_{p-1,k},\dots,y_{p-Q_k+1,k}]^{\top} \nonumber \\
 &\mathbf{g}_k=\left[\frac{b_{0,k}}{a_{0,k}},\dots,\frac{b_{Q_k-1,k}}{a_{0,k}},-\frac{a_{1,k}}{a_{0,k}},\dots,-\frac{a_{Q_k-1,k}}{a_{0,k}}\right]^{\top}.
\end{align}
We see that the DP-RTF appears as the first entry of $\mathbf{g}_k$. Hence, in the following, we base the estimation of the DP-RTF on the construction of $y_{p,k}$ and $\mathbf{z}_{p,k}$ statistics.
More specifically, multiplying both sides of (\ref{zpk}) by $ y_{p,k}^*$ ($^*$ denotes complex conjugation) and taking the expectation (denoted by $E\{\cdot\}$), we obtain:
 \begin{align}\label{phi}
\phi_{yy}(p,k) = \varphi_{zy}^{\top}(p,k) \mathbf{g}_k,
 \end{align}
where $\phi_{yy}(p,k)=E\{y_{p,k}y_{p,k}^{*}\}$ is the PSD of $y(p,k)$, and
\begin{align}
 \varphi_{zy}(p,k)& = [E\{x_{p,k}y_{p,k}^*\},\dots,E\{x_{p-Q_k+1,k}y_{p,k}^*\}, \nonumber \\
 &E\{y_{p-1,k}y_{p,k}^*\},\dots,E\{y_{p-Q_k+1,k}y_{p,k}^*\}]^{\top}
\end{align}
is a vector which is composed of cross-PSD terms between the elements of $\mathbf{z}_{p,k}$ and $y_{p,k}$.
In practice, these auto- and cross-PSD terms can be estimated by averaging the corresponding spectra over a number $D$ of frames, i.e.:
 \begin{align}\label{hphi}
\hat{\phi}_{yy}(p,k) = \frac{1}{D}\sum_{d=0}^{D-1}y_{p-d,k}y_{p-d,k}
 \end{align}
The elements in $\varphi_{zy}(p,k)$ can be estimated by using the same principle. Consequently, (\ref{phi}) is approximated as
 \begin{align}\label{hatphi}
 \hat{\phi}_{yy}(p,k) = \hat{\varphi}_{zy}^{\top}(p,k) \mathbf{g}_k.
 \end{align} 
In this equation, the speech PSD $\hat{\phi}_{yy}(p,k)$ and $\hat{\varphi}_{zy}^{\top}(p,k)$ can be obtained from the noise-free sensor signals. However in the real world, the PSD of speech signals are deteriorated by noise.  

\subsection{Speech PSD Estimate in the Presence of Noise}
\label{sec32}

Noise signals are added into the sensor signals in (\ref{xn}) as
\begin{align}\label{xnu}
 \tilde{x}(n)=x(n)+u(n)=a(n)*s(n)+u(n), \nonumber \\
 \tilde{y}(n)=y(n)+v(n)=b(n)*s(n)+v(n), 
\end{align}
where $u(n)$ and $v(n)$ are the noise signals in two sensors, respectively, which are supposed to be stationary and uncorrelated to the speech signal $s(n)$.    

Applying the STFT to the sensor signals in (\ref{xnu}): $\tilde{x}_{p,k} = x_{p,k}+u_{p,k}$ and $\tilde{y}_{p,k} = y_{p,k}+v_{p,k}$, respectively, in which each quantity is the STFT coefficient of its corresponding time domain signal.
Similar to ${\mathbf{z}}_{p,k}$, we define
\begin{align} 
 \tilde{\mathbf{z}}_{p,k} &= [\tilde{x}_{p,k},\dots,\tilde{x}_{p-Q_k+1,k},\tilde{y}_{p-1,k},\dots,\tilde{y}_{p-Q_k+1,k}]^{\top} \nonumber \\
 & =\mathbf{z}_{p,k}+\mathbf{w}_{p,k} 
\end{align}
where 
\begin{align}
 \mathbf{w}_{p,k}=[u_{p,k},\dots,u_{p-Q_k+1,k},v_{p-1,k},\dots,v_{p-Q_k+1,k}]^{\top}.
\end{align}

We define the PSD of $\tilde{y}_{p,k}$ as $\phi_{\tilde{y}\tilde{y}}(p,k)$. We also define the PSD vector $\varphi_{\tilde{z}\tilde{y}}(p,k)$, which is composed of the auto- or cross-PSD between the elements of $\tilde{\mathbf{z}}_{p,k}$ and $\tilde{y}_{p,k}$.
Following the principle in (\ref{hphi}), by averaging the auto or cross spectra of multiple frames, these PSDs can be estimated using the STFT coefficients of input signals as $\hat{\phi}_{\tilde{y}\tilde{y}}(p,k)$ and $\hat{\varphi}_{\tilde{z}\tilde{y}}(p,k)$.
Because the speech and noise signals are not correlated, they can be represented as
\begin{align}\label{hatphin}
 \hat{\phi}_{\tilde{y}\tilde{y}}(p,k) = \hat{\phi}_{yy}(p,k)+\hat{\phi}_{vv}(p,k) \nonumber \\
 \hat{\varphi}_{\tilde{z}\tilde{y}}(p,k) = \hat{\varphi}_{zy}(p,k)+\hat{\varphi}_{wv}(p,k)
\end{align}
where $\hat{\phi}_{vv}(p,k)$ is an estimation of the PSD of $v_{p,k}$, the PSD vector $\hat{\varphi}_{wv}(p,k)$ is composed of the estimated auto or cross PSD  between the elements of ${\mathbf{w}}_{p,k}$ and ${v}_{p,k}$. 
The auto- and cross-PSD of noise can be subtracted by using the noise estimator \cite{cohen2001} or the inter-frame spectral subtraction technique \cite{mine2015assp}.   
In this work, for simplicity, we assume that noise is stationary (for example, the robot's ego-noise), and the noise-only signal is available, from which the noise PSD ${\phi}_{vv}(p,k)$ and ${\varphi}_{wv}(p,k)$ can be computed in advance.
Consequently, we approximately compute the speech PSD as
\begin{align}\label{ssphi}
 \hat{\phi}_{yy}(p,k) & \approx  \hat{\phi}_{\tilde{y}\tilde{y}}(p,k) - {\phi}_{vv}(p,k) \nonumber \\
 \hat{\varphi}_{zy}(p,k) & \approx \hat{\varphi}_{\tilde{z}\tilde{y}}(p,k) -{\varphi}_{wv}(p,k).	 
\end{align}
Because of the temporal sparsity of the speech signal, parts of the frames are dominated by noise, which should be disregarded for DP-RTF estimation.
Thence we define the frame index set  that comprises the frames with considerable speech power as 
\begin{align}
\mathbf{p}_k=\{p\ | \ \hat{\phi}_{\tilde{y}\tilde{y}}(p,k)> \gamma{\phi}_{vv}(p,k)\},
\end{align}
where $\gamma$ is a power threshold. Let $P_k=|\mathbf{p}_k|$ denote the cardinal of $\mathbf{p}_k$.

\subsection{Direct-Path Relative Transfer Function Estimation}

Based on the speech PSD estimated in (\ref{ssphi}), by concatenating across frames, (\ref{hatphi}) could be written in matrix form
 \begin{align}\label{Phi}
 \hat{\Phi}_{yy}(k) = \hat{\Psi}_{zy}(k) \mathbf{g}_k.
 \end{align}
where 
\begin{align}
 &\hat{\Phi}_{yy}(k)=[\dots,\hat{\phi}_{yy}(p,k),\dots]^{\top},  \quad p\in \mathbf{p}_k  \nonumber \\
 &\hat{\Psi}_{zy}(k)=[\dots,\hat{\varphi}_{zy}(p,k),\dots]^{\top}, \quad  p\in \mathbf{p}_k \nonumber
\end{align}
are $P_k\times1$ vector, $P_k\times(2Q_k-1)$ matrix, respectively.

A least-square (LS) solution to (\ref{Phi}) is given as
\begin{align}
 \hat{\mathbf{g}}_k = (\hat{\Psi}_{zy}^H(k)\hat{\Psi}_{zy}(k))^{-1}\hat{\Psi}_{zy}(k)\hat{\Phi}_{yy}(k)
\end{align}
where $^H$ denotes matrix conjugate transpose, $^{-1}$ denotes maxtrix inverse. The first element of $\hat{\mathbf{g}}_k$ is denoted as $\hat{g}_{k}$, which is an estimation of DP-RTF $\frac{b_{0,k}}{a_{0,k}}$.

\section{Sound Source Localization Method}
\label{secssl}
The amplitude and the phase of DP-RTF is equivalent to the  IPD and ILD interaural cues corresponding to the direct-path propagation. 
As discussed in \cite{araki2007,mine2015eusipco}, when the reference transfer function $a_{0,k}$  is much smaller than $b_{0,k}$, 
the amplitude ratio estimation is sensitive to the noise in the reference channel. Therefore, we normalize $\hat{g}_{k}$ as 
\begin{align}
 \hat{c}_k = \frac{\hat{g}_{k}}{\sqrt{|\hat{g}_{k}|^2+1}}.
\end{align}
It is clear that the phase is retained, and the amplitude is normalized as $0<|\hat{c}_k|<1$. 

The quantity $\hat{c}_k$ is the estimated DP-RTF for one microphone pair, where the index of microphone pair is omitted.
Concatenating the estimated DP-RTF of microphone pairs A-C and B-D, yields $\hat{\mathbf{c}}_k=[\hat{c}_{k,AC},\hat{c}_{k,BD}]^{\top}$.\footnote{For NAO version~5, a total of six microphone pairs are available. However, experiments show that it is sufficient to consider two microphone pairs.}
Then, concatenating $\hat{\mathbf{c}}_k$ across frequencies, we obtain a global feature vector in ${\mathbb{C}}^{2K}$: 
\begin{equation}
\label{eq:global-vector}
\hat{\mathbf{c}}=[\hat{\mathbf{c}}_0^{\top},\dots,\hat{\mathbf{c}}_k^{\top},\dots,\hat{\mathbf{c}}_{K-1}^{\top}]^{\top},
\end{equation}
where $K$ denotes the number of frequencies involved in SSL. 

To map the high-dimensional feature vector $\hat{\mathbf{c}}$ to a low-dimensional source direction $\mathbf{o}\in\mathbb{R}^O$ ($O$ denote the dimension of source direction), we adopt the regression method proposed in \cite{deleforge2015b}.
Briefly, a probabilistic piecewise-linear regression $f:\mathbb{C}^{2K}\rightarrow \mathbb{R}^O$ is learned from a training dataset $\{\mathbf{c}_i, \mathbf{o}_i\}_{i=1}^I$, where $\mathbf{c}_i$ is a feature vector  and $\mathbf{o}_i$ is the corresponding sound-source direction. 
Then, for a test DP-RTF feature vector $\hat{\mathbf{c}}$ extracted from the microphone signals, the source direction is predicted with $\hat{\mathbf{o}}=f(\hat{\mathbf{c}})$.
 
Due to the sparsity of speech signals in the STFT domain, it is possible that there are only a few significant speech frames at frequency $k$ for one microphone pair, especially in the case of low SNR. 
In other words, $P_{k}$ could be small, which makes the estimated $\hat{c}_{k}$ unreliable. To disregard the unreliable $\hat{c}_{k}$ in the regression procedure, 
we introduce a missing data indicator vector $\mathbf{h}\in\mathbb{R}^{2K}$. If the matrix $\hat{\Psi}_{zy}(k)$ in (\ref{Phi}) is underdetermined, i.e., $P_{k}<2Q_k-1$, 
its corresponding element in $\mathbf{h}$ is set to 0, and 1 otherwise. The regression method that we use \cite{deleforge2015b} makes use of such an indicator vector $\mathbf{h}$ and the element in $\hat{\mathbf{c}}$ with a 0 indicator is disregarded.
The revised prediction is $\hat{\mathbf{o}}=f(\hat{\mathbf{c}},\mathbf{h})$.

\section{Experiments with the Nao Robot}
\label{secexp}

In this section several experiments using the NAO robot (version 5) are conducted in various real-world environments. From Fig.~\ref{fignao}, 
one can see that four microphones are nearly coplanar, and that the angle between the microphone plane and the horizontal plane is small. 
The microphones are close to the head's fan (the circular ear in Fig.~\ref{fignao}), thence the microphone recording include ego-noise due to the fan.
As mentioned in \cite{loellmann2014}, the fan noise is stationary and spatially correlated. 
In addition, its spectral energy mainly concentrates in a frequency range of up to 4 kHz, thence the recorded speech signal will be contaminated by the fan noise significantly. 

\subsection{The Datasets}
\begin{figure}[t]
\centering
{\includegraphics[height=0.55\columnwidth]{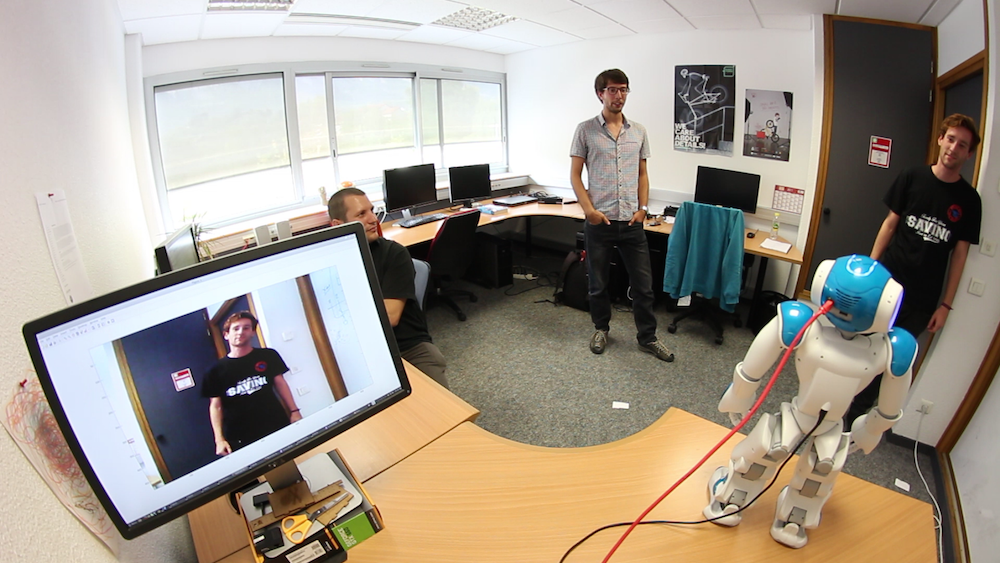}}
\caption{A typical \textit{audio-only} localization experiment in the office environment. The robot turns its head towards the speaking person shown on the screen (please see the supplementary video).} 
\label{fig:office-ssl}
\end{figure}

The data are recorded in four real world environments: meeting room, laboratory, office, e.g., Fig.~\ref{fig:office-ssl}, and cafeteria, whose reverberation time $T_{60}$ are approximately 1.04 s, 0.52 s, 0.47 s and 0.24 s, respectively.

\begin{figure}[t]
\centering
{\includegraphics[height=0.55\columnwidth]{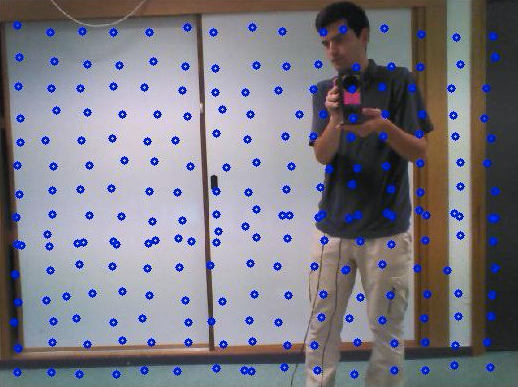}}
\caption{The \textit{audio-visual} training dataset contains sound sources emitted by a loud-speaker that correspond to sound directions materialized by image locations (marked as blue circles).} 
\label{figgrid}
\end{figure}

Two \textbf{test datasets} are recorded in these environments: 
\begin{itemize}
\item The \emph{Audio-only} dataset: in the laboratory, the speech recording from the TIMIT dataset \cite{garofolo1988} are emitted by a loudspeaker. 
Two groups of data are recorded with a fixed robot-to-source distance of 1.1 m and 2.5 m, respectively. Besides $T_{60}$, ITDG and direct-to-reverberation ratio (DRR) are also important to measure the intensity of the reverberation. In general, the larger the robot-to-source distance the less ITDG and DRR. 
Obviously, the two cross microphone pairs allow a $360^\circ$ azimuth localization. However, because of the limitation of NAO's head joint, NAO's head can not rotate in a $360^\circ$ azimuth range. 
Thence, for each group, 174 sounds are emitted from directions uniformly distributed in the range $-120^\circ$ to $120^\circ$ (azimuth), and $-15^\circ$ to 25$^\circ$ (elevation).
\item The \emph{Audio-visual} dataset: Fig. \ref{fignao} shows the NAO head camera, with a field-of-view  of 60.97$^\circ\times$47.64$^\circ$; speech sounds are emitted by a loudspeaker lying in the camera's field of view. 
The image resolution is of 640$\times$480 pixels, so 1$^\circ$ of azimuth/elevation corresponds to about 10.5 horizontal/vertical pixels. For this dataset, the source direction corresponds to a pixel in the image. 
The ground-truth source direction is obtained by localizing in the image the visual marker fixed on the loudspeaker. 
Four groups of data are recorded in four rooms, respectively. For each group, about 230 sounds are emitted from directions uniformly distributed in the the camera field-of-view.
As an example, Fig.~\ref{figgrid} illustrates the 228 directions shown as blue dots in the image plane. The robot-to-source distance is approximately fixed as 2 m in this dataset.
\end{itemize}


In both of these two datasets, the external noise is much lower than the fan noise, thence noise in the recorded signal is almost composed of the fan noise. 
The signal to noise ratios (SNR) are approximately 14~dB, 11~dB for \emph{Audio-only} dataset with 1.1~m and 2.5~m robot-to-source distance, respectively, and 2~dB for \emph{audio-visual} dataset \footnote{Note that the loudspeaker volume is different for two datasets.}. 
As mentioned in Section~\ref{sec32}, the fan noise PSD ${\phi}_{vv}(p,k)$ and ${\varphi}_{wv}(p,k)$ are precomputed.

The \textbf{training dataset} $\{\mathbf{c}_i, \mathbf{o}_i\}_{i=1}^I$ for \emph{Audio-only} experiments is generated by the anechoic head-related impulse responses (HRIR) of 1002 directions uniformly distributed in 
the same range as the test dataset. The training dataset for \emph{Audio-visual} experiments is generated by the HRIR of 378 directions uniformly distributed in the camera field-of-view.
The anechoic HRIR is obtained by truncating the room impulse response before the first reflection. White Gaussian noise (WGN) signals are emitted from each direction, and the cross-correlation 
between the microphone signal and source WGN signal gives the room impulse response of each direction.


\subsection{Parameter Setup}

The sampling rate of the microphone signals is 16 kHz. The window length of STFT is 16 ms (256 samples) with 8 ms  overlap (128 samples). Only the frequency band from 300~Hz to 4~kHz is taken into account for speech source localization, 
i.e., the corresponding frequency bins are from 5 to 63, so the number of frequencies is $K$=59. The number of frames $D$ for PSD estimation is set to 25 (0.2 s). The power threshold $\gamma$ is set to 1.8. 
We set the length of CTF $Q_k$ to be equal for all the frequency bins for simplicity, and denote it as $Q$, which is set to $0.25T_{60}$.  

\subsection{Method Comparison}
The crucial point of binaural localization is to extract the reliable binaural cues from the noisy and reverberant sensor signals. 
Two state-of-the-art binaural feature estimation methods with good capability to reduce noise or reverberations are tested for comparison. 
\begin{itemize}
\item A variation of the unbiased RTF estimator proposed in \cite{cohen2004}, in which the MTF approximation is adopted. 
The noise PSD is recursively estimated in the original work, while is more accurately precomputed using the noise-only signal in this work.  We refer to this method as RTF-MTF. 
\item The coherence test (CT) method in \cite{mohan2008}. The coherence test is used for searching the rank-1 time-frequency bins, which are supposed to be dominated by one active source. 
In this work, it is adopted for single speaker localization, in which one active source denotes the direct-path source signal. The TF bins that involve considerable reflections have low coherence.
We first detect the maximum coherence over all the frames at each frequency bin, and then set the coherence test threshold for each frequency bin to 0.9 times its maximum coherence. In our experiments, this threshold achieves the best performance.
The covariance matrix is estimated by taking a 120 ms (15 adjacent frames) averaging. The auto and cross PSD of all the frames that have a coherence larger than the threshold are applied the spectral subtraction with the same principle 
in (\ref{ssphi}), and then are averaged over frames for acoustic feature extraction. We refer to this method as RTF-CT. 
\item In addition, a conventional beamforming SSL method: the steered-response power (SRP) utilizing the phase transform (PHAT) \cite{dibiase2001,do2007} is also tested. 
The source directions in the training set of the proposed method are taken as the steering directions, and their HRIRs are taken as the steering vector.
\end{itemize}

\subsection{Localization Results with the Audio-Only Dataset}

\begin{figure}[t]
\centering
\subfloat[]{\includegraphics[width=0.45\textwidth]{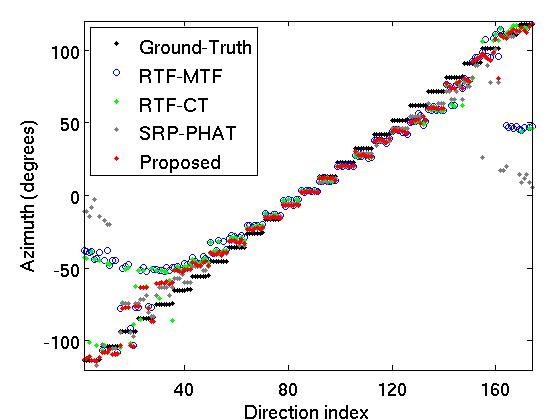}} \\ \vspace{-0.3cm}
\subfloat[]{\includegraphics[width=0.45\textwidth]{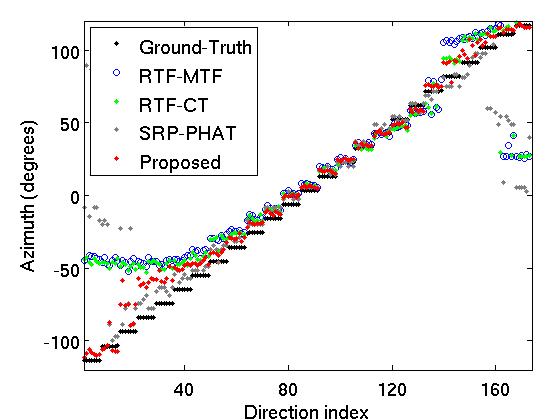}}
\caption{Localization results for \emph{Audio-only} dataset. (a) 1.1 m robot-to-source distance. (b) 2.5 m robot-to-source distance. The elevations of multiple source directions corresponding to each azimuth uniformly distribute from -15$^\circ$ to 25$^\circ$.} 
\label{figres}
\end{figure}

Our experiments on \emph{Audio-only} dataset show that, in the elevation range $[-15^\circ\, 25^\circ]$, the elevation localization results are completely unreliable for all the three methods. This can be easily explained by the fact that the angle between the microphone plane and the horizontal plane is small, hence the microphone array has a low resolution for the elevation direction. 
Therefore, in Fig. \ref{figres}, we only present the azimuth localization results.

From Fig.~\ref{figres}-(a), we observe that both the proposed method and the RTF-MTF and RTF-CT methods work well in the azimuth range $[-50^\circ,\ 50^\circ]$. 
The proposed method achieves slightly better results in this range. The performance drops drastically for the source directions out of this range. 
This indicates that the NAO's microphone array has a better localization capability for the azimuth range $[-50^\circ,\ 50^\circ]$.
From the results for the azimuth range $[-120^\circ,\ -50^\circ]$ and $[50^\circ,\ 120^\circ]$, it can be seen that RTF-MTF has the largest localization error and many localization outliers 
caused by the reverberations. By selecting frames that involve less reverberations, RTF-CT performs better than RTF-MTF, evidently, which can be observed from the fact that RTF-CT has less outliers than RTF-MTF.
However, it is difficult to automatically set a coherence test threshold that could perfectly select the desired frames. Many frames that have a coherence larger than the threshold include reflections.
Therefore, RTF-CT also has a relatively large localization error and some localization outliers. There are also many outliers for SRP-PHAT, which indicates that the steered response power is influenced by the reverberation.
The proposed method achieves the best performance by properly extracting the direct-path RTF. 

Fig. \ref{figres}-(b) shows the localization results for the data with 2.5~m robot-to-source distance. 
Compared to the robot-to-source distance of 1.1~m, both ITDG and DRR are smaller. Consequently, the performance degrades for both the proposed method and the two state-of-the-art methods 
compared to Fig.~\ref{figres}-(a). The reasons for this degradation are the followings: for both RTF-MTF and RTF-CT the reflections are large relative to the direct-path impulse response, which makes the feature estimated from the reverberated signals more different than the feature corresponding to the direct-path propagation.
In addition, concerning RTF-CT, the early reflection is closer to the direct-path impulse response, which makes less reverberation-free TF bins to be available. 
SRP-PHAT also has more outliers than the case in Fig.~\ref{figres}-(a) due to the lower DRR.
For the proposed method, (i)~the early reflections in the impulse response segment $a(n)|_{n=0}^N$ increase and (ii) in vector $\mathbf{g}_k$, the DP-RTF $\frac{b_{0,k}}{a_{0,k}}$ plays a more unimportant role relative to the other elements with the decreasing of DRR, which makes the DP-RTF estimation error larger.
We can see that the proposed method still achieves the best performance, and most of its localization results are reliable.

\subsection{Localization Results with the Audio-Visual Dataset}

The source directions of \emph{audio-visual} dataset distribute in the camera field-of-view, which is a small range in front of NAO's head (azimuth range $[-30.5^\circ,\ 30.5^\circ]$). As shown in Fig. \ref{figres}, good azimuth localization results
are obtained in this range. Table \ref{tle} shows the localization error for both the azimuth (Azi.) and elevation (Ele.) directions. The localization error is computed by averaging all the absolute errors between the localized directions and their corresponding ground truth (in degrees).
\begin{figure*}[t!h!]
\centering
{\includegraphics[width=0.85\textwidth]{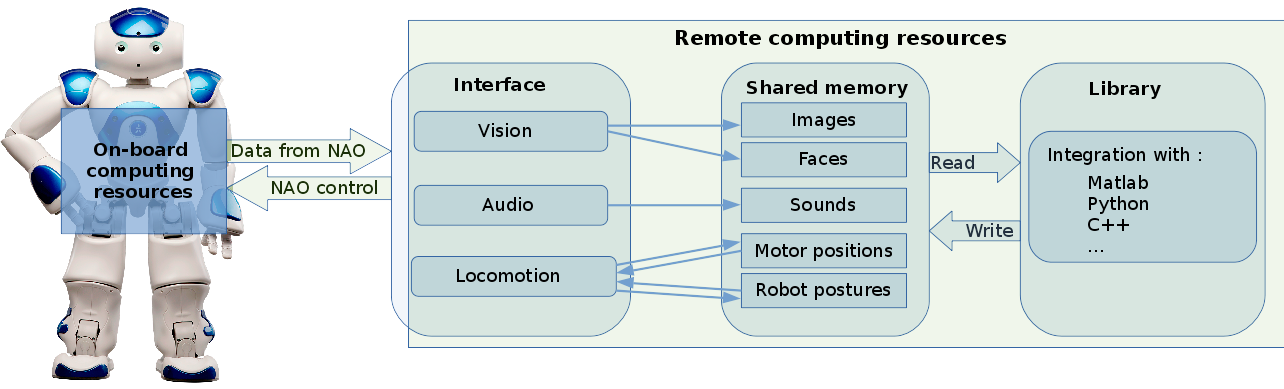}}
\caption{Overview of the proposed distributed architecture that allows fast development of interactive applications using the humanoid robot NAO \cite{badeig2015distributed}.} 
\label{fig:software}
\end{figure*}

It can be seen that the elevation errors are always much bigger than the azimuth errors, due to the low elevation resolution of the microphone array that we already mentioned.
In the cafeteria, the reverberation time $T_{60}$ is 0.24 s, generally speaking, which is a low reverberation time. 
The RTF-MTF and RTF-CT methods yields performance comparable with the proposed method in the cafeteria environment. 
The reason is: the MTF approximation is relatively proper for this case, while the proposed method has a higher model complexity which needs more reliable data.
In the office and laboratory, the reverberation times are larger, so the MTF approximation is not accurate anymore. 
As a result, Table \ref{tle} shows that the proposed method achieves evidently better performance than the two other methods in the office and laboratory environments. 
The performance of RTF-MTF is even better than RTF-CT, the reason is probably that the coherence test doesn't work well under low SNR conditions (the SNR is about 2 dB).
In the meeting room, the reverberation time is high (1.04 s). 
SRP-PHAT achieves the worst performance due to the intense noise, especially the noise is spatially correlated.
The proposed method still evidently performs better than the other methods.  
These further validates that the proposed method is more efficient in reverberant environments. 

\setlength{\tabcolsep}{5.2pt}
\begin{table}[h!]
\centering
\begin{tabular}{| c | c  c | c  c | c  c |c c|}
\hline         
	    & \multicolumn{2}{c|}{Cafeteria} & \multicolumn{2}{c|}{Office} & \multicolumn{2}{c|}{Laboratory} & \multicolumn{2}{c|}{Meeting room}  \\
 Methods    & Azi.      & Ele.      & Azi.      & Ele.     & Azi.     & Ele.    & Azi.     & Ele.                               \\ \hline
 RTF-MTF        & 0.45      & 1.57      & 0.62     & 2.14     & 1.44   & 2.31      &    1.87 & 3.66                            \\
 RTF-CT         & \textbf{0.44}  & 1.50      & 0.64     & 2.25     & 1.61   & 2.36    & 1.77 & 3.44      \\ 
 SRP-PHAT        & 0.77      & 1.95      & 1.03     &2.80          & 1.41  & 3.33     & 2.04 & 3.52                                \\
 Proposed   & 0.47      & \textbf{1.47}      & \textbf{0.55}     & \textbf{1.87}     & \textbf{0.82}   & \textbf{1.84}  & \textbf{0.95}   & \textbf{2.12}       \\ \hline
\end{tabular}
\caption{Localization error (degrees) for the \emph{audio-visual} dataset. The best results are shown in \textbf{bold}.}
\label{tle}
\vspace{-.3cm}
\end{table}

\subsection{Software Architecture}

Ideally, one would like to implement the SSL method just presented using the embedded computing resources available with a robot such as the NAO companion humanoid.
However, NAO like any other commercially available robot, has two limitations. Firstly, the on-board computing resources are restricted which implies that it is difficult to implement sophisticated audio signal processing and analysis algorithms needed by SSL in particular and by robot audition in general. Secondly, robot programming implies the development of embedded software modules and libraries, which is a difficult task in its own right necessitating specialized knowledge. 

We have developed a distributed software architecture that attempts to overcome these two limitations and which allows fast experimental validation of proof-of-concept demonstrators \cite{badeig2015distributed}. Broadly speaking, NAO's on-board computing resources are networked with external (or remote) computing resources. The latter is a computer platform (laptop or desktop) with its CPU's, GPU's, memory, operating system, libraries, software packages, internet access, etc. This configuration enables easy and fast development in Matlab, C, C++, Python, etc. Moreover, it allows the user to combine on-board libraries (motion control, face detection, etc.) with external toolboxes, such as Matlab's signal processing toolbox.

An overview of the proposed software architecture is shown on Fig.~\ref{fig:software}. Data coming from NAO (motor positions, images, microphone signals, or data produced by on-board computing modules) are fed into the external computer. Conversely, the latter can control the robot. Currently we developed three internal-to-remote interfaces: \textit{vision}, \textit{audio}, and \textit{locomotion}. The role of these interfaces  is twofold: (i)~to feed the data into a memory space that is subsequently shared with existing software modules or with modules under development and (ii)~to send back to the robot commands generated by the external software modules. Although these modules may be developed in a variety of programming languages, special emphasis was put to allow integration with the Matlab programming environment. 

The proposed SSL method is implemented in Matlab, which offers the possibility to use Matlab's signal processing toolbox, e.g., the STFT. The Matlab computer vision toolbox is used for image processing.  The on-board robot controller is invoked to rotate the robot head in the direction of the detected sound source.
\section{Conclusions}
\label{seccon}

We have proposed a direct-path RTF estimator for SSL, and tested it on NAO robot. Instead of the MTF approximation, the method takes the CTF approximation, 
which is more precise when the impulse response is too long. Compared with the conventional RTF, the ratio between two direct-path ATFs is more reliable for SSL. 
Because the trainning dataset is generated using the anechoic HRIR, the SSL module can operate for various room configurations, which is important for robot audition. 
Experiments have shown that the proposed method performs well for azimuth localization under difficult acoustic conditions, however poorly for elevation localization because of the microphone geometry of NAO robot head version~5.  
Thence, for the next version of NAO, a more reasonable microphone topology is expected. 




\end{document}